\documentclass[11pt]{article}
\usepackage{moriond,epsfig}

\usepackage{amsmath}
\usepackage{amsfonts}
\usepackage{amssymb}

\def\be{\begin{equation}}
\def\ee{\end{equation}}
\def\bea{\begin{eqnarray}}
\def\eea{\end{eqnarray}}

\def\tev{\, {\rm TeV}}

\newcommand{\tr}{\mathrm T \mathrm r}

\newcommand{\nn}{\nonumber \\}

\newcommand{\Photo}{\includegraphics[height=35mm]{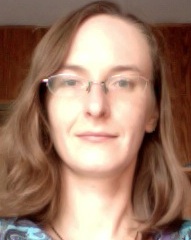}}

\begin{document}
\vspace*{4cm}
\title{COMPOSITE OR ELEMENTARY? PROBING THE NATURE OF THE HIGGS}

\author{ANNA KAMI\'{N}SKA}

\address{Rudolf Peierls Centre for Theoretical Physics, University of Oxford, 1 Keble Road, OX1 3NP, Oxford, UK}

\maketitle\abstracts{
I discuss consequences of electroweak symmetry breaking by strong dynamics, assuming the existence of a light composite scalar appearing as a pseudo-Goldstone boson of some global symmetry of the new strongly interacting sector. In such a scenario, the composite scalar has properties very similar to the Standard Model Higgs, but the existence of additional resonances with different spins is also expected. Properties and phenomenology of lightest spin-1 resonances are considered in a simple general effective Lagrangian description. The question whether the effects of spin-1 resonances can be observed at the LHC, shedding light on the nature of the Higgs boson, is addressed.}

\section{Introduction}

A composite Higgs boson originating from electroweak symmetry breaking induced by some new strongly interacting sector is a viable alterative to an elementary, Standard Model (SM) - like Higgs scenario. It does not suffer under the hierarchy problem and its properties at energies low with respect to the scale of strong dynamics can be very similar to the properties of the SM scalar. Moreover, the composite Higgs boson can be made naturally light with respect to the compositeness scale when it appears as a pseudo-Goldstone boson (PGB) of the global symmetry breaking of the strong sector. The minimal model of this kind is based on the $SO(5)\rightarrow SO(4)$ global symmetry breaking pattern \cite{Agashe:2004rs}, which produces a full pseudo-Goldstone composite Higgs doublet.

In order to gain a better understanding of the electroweak symmetry breaking mechanism it is crucial to look for signatures distinguishing a composite Higgs boson from an elementary Higgs boson. First of all, the couplings of a PGB composite Higgs to $W$ and $Z$ gauge bosons and to fermions are modified with respect to the couplings of an elementary Higgs boson. The modification is scaled by the parameter $\xi=(v/f_{\pi})^2$, which describes the squared ratio of the electroweak scale to the scale of strong dynamics. In the minimal composite Higgs model the couplings of the composite scalar to $W$ and $Z$ gauge bosons is given by
\begin{equation}
\mathcal{L}^{\left( 2\right) } =\frac{1}{2}\partial_{\mu}h\partial^{\mu}h+\frac{v^2}{4}\left( 1+2a\frac{h}{v}+b\frac{h^2}{v^2}+...\right) \tr\left\lbrace D^{\mu}U\; D_{\mu}U^{\dag} \right\rbrace 
\end{equation}
where $U=e^{i\pi^a(x)\sigma^a/v}$ parameterizes Goldstone bosons corresponding to longitudinal $W$ and $Z$ and transforms as $g_{L}Ug_{R}^{\dag}$ under the global $SU(2)_{L}\times SU(2)_{R}$ group gauged by the electroweak group. Couplings $a= \sqrt{1-\xi},\ b= 1- 2\xi$ reproduce the Standard Model values $a_{SM}=b_{SM}=1$ in the limit $\xi\rightarrow 0$ with $v=const$. Hence, we expect $\xi$ to be small, $\xi\lesssim 0.3$. In fact, the departure from SM-like Higgs couplings might not be detectable at the LHC. Hence it is worth to look simultaneously for other signatures of strong electroweak breaking related to resonances of different spins. Both spin-1/2 and spin-1 resonances should contribute to indirect (electroweak precision tests, flavor observables) and direct signals of physics beyond the Standard Model. Here I focus on effects of spin-1 resonances, constructing a simple, general and self-consistent effective framework to study their properties and LHC phenomenology.

\section{Building the effective Lagrangian}

In order to introduce spin-1 resonances into the effective description I use the formalism of hidden local symmetry \cite{Bando:1987br}, which works very well for the description of vector resonances in low energy QCD. Generalizing this formalism to arbitrary global symmetry breaking patterns $\mathcal{G}\rightarrow\mathcal{H}$ means that in order to write down the effective Lagrangian for the analogue of vector resonances one has to add a new 'hidden' gauge group $\mathcal{H}_{local}$, which gauge bosons provide the degrees of freedom needed for the description of spin-1 resonances. Then the new building block for the effective Lagrangian is a sigma-model field $S$ related to $\mathcal{G}\times\mathcal{H}_{local}\rightarrow\mathcal{H}$ symmetry breaking and transforming as
\begin{equation}
S\ \rightarrow\ g\; S\; h^{\dag},\ \ \ \ \ \ g\in\mathcal{G},\ h\in\mathcal{H}_{local},\ \ \ \left\langle S\right\rangle =\textbf{1}.
\end{equation}
The leading order effective Lagrangian term
\begin{equation}
\mathcal{L} \ni v_{1}^{2}\tr\left\lbrace D_{\mu}SD^{\mu}S^{\dag}\right\rbrace
\end{equation}
introduces through covariant derivatives of $S$ mixing terms and interactions between gauge fields of the Standard Model electroweak group sitting in $\mathcal{G}$ and the new 'hidden' local group. The symmetry breaking pattern can be further modified to take into account more and more spin-1 states. Eigenstates in the spin-1 sector are now combinations of $W^a_{\mu}$, $B_{\mu}$ and hidden gauge $\rho_{\mu}$ fields. However, if we assume a hierarchy of couplings $g,g'\ll g_{\rho}$, the lightest eigenstates correspond at the leading order in $g/g_{\rho}$ to Standard Model $W$, $Z$ and photon fields, while the heavy eigenstates correspond at the leading order to $\rho$ fields. Moreover, for small values of $\xi$, there exists a 'pairing up' between gauge bosons corresponding to $SU(2)_{L}$ ($SU(2)_{R}$) and $SU(2)_{L\; hidden}$ ($SU(2)_{R\; hidden}$) groups respectively, which leads to the conclusion that $\rho_{L}$ and $\rho_{R}$ resonances have the strongest interactions with Standard Model particles and are most important from the point of view of LHC phenomenology. 

\section{LHC phenomenology}

\begin{figure}[h!]
\begin{minipage}[b]{0.47\linewidth}
\centering
\includegraphics[width=7.5 cm]{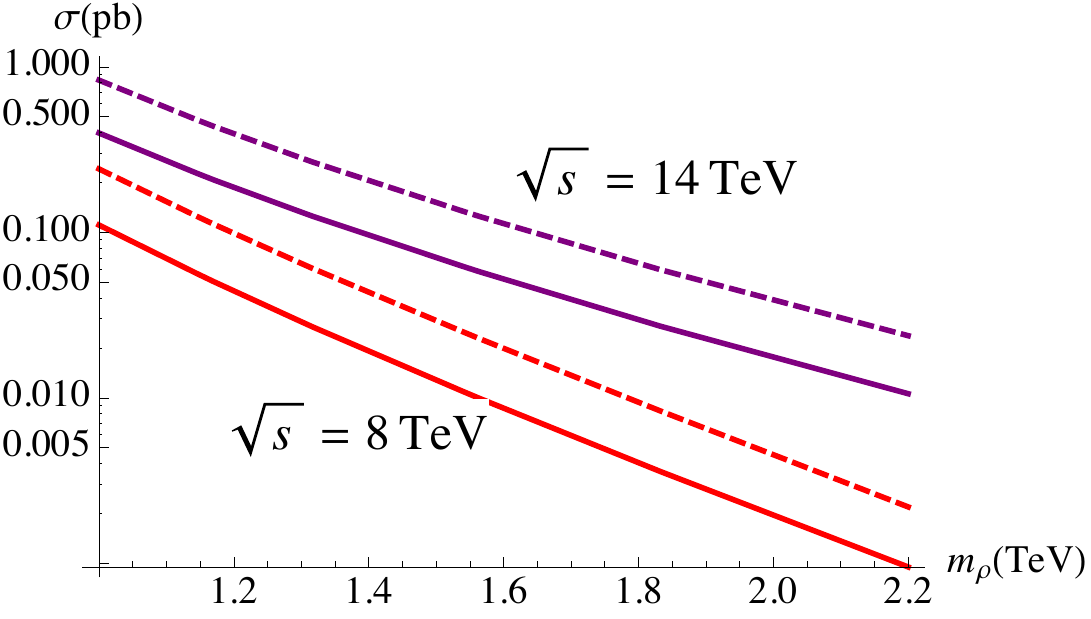}
\caption{Drell-Yan production cross section of $\rho_{L}$.}
\label{DYg4SO5xi02}
\end{minipage}
\hspace{0.5cm}
\begin{minipage}[b]{0.47\linewidth}
\centering
\includegraphics[width=6.9 cm]{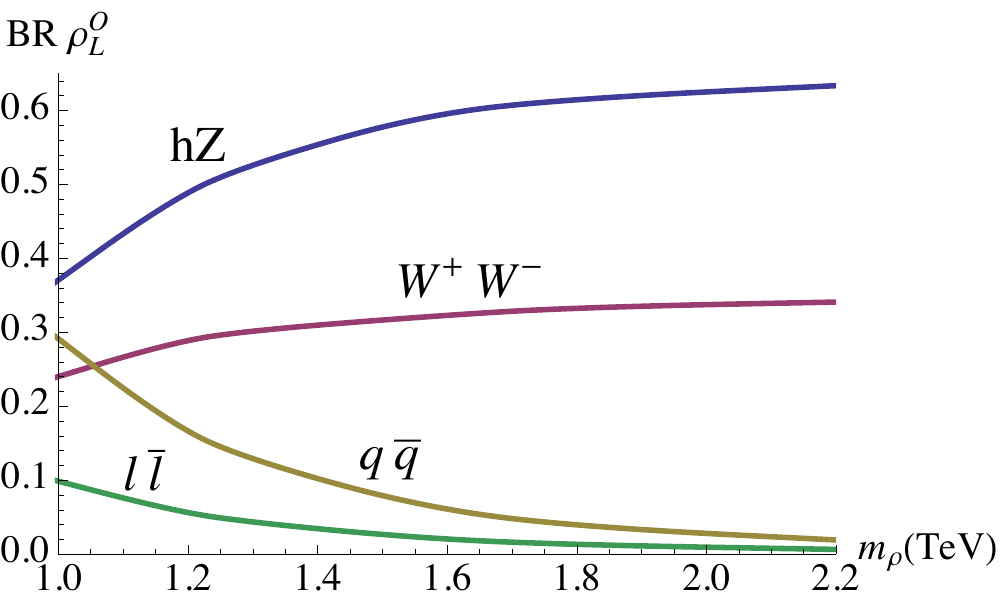}
\caption{Branching ratios of a neutral $\rho_{L}$.}
\label{BRv0g4SO5xi02}
\end{minipage}
\end{figure}

\begin{figure}[h!]
\begin{minipage}[b]{0.47\linewidth}
\centering
\includegraphics[width=70 mm]{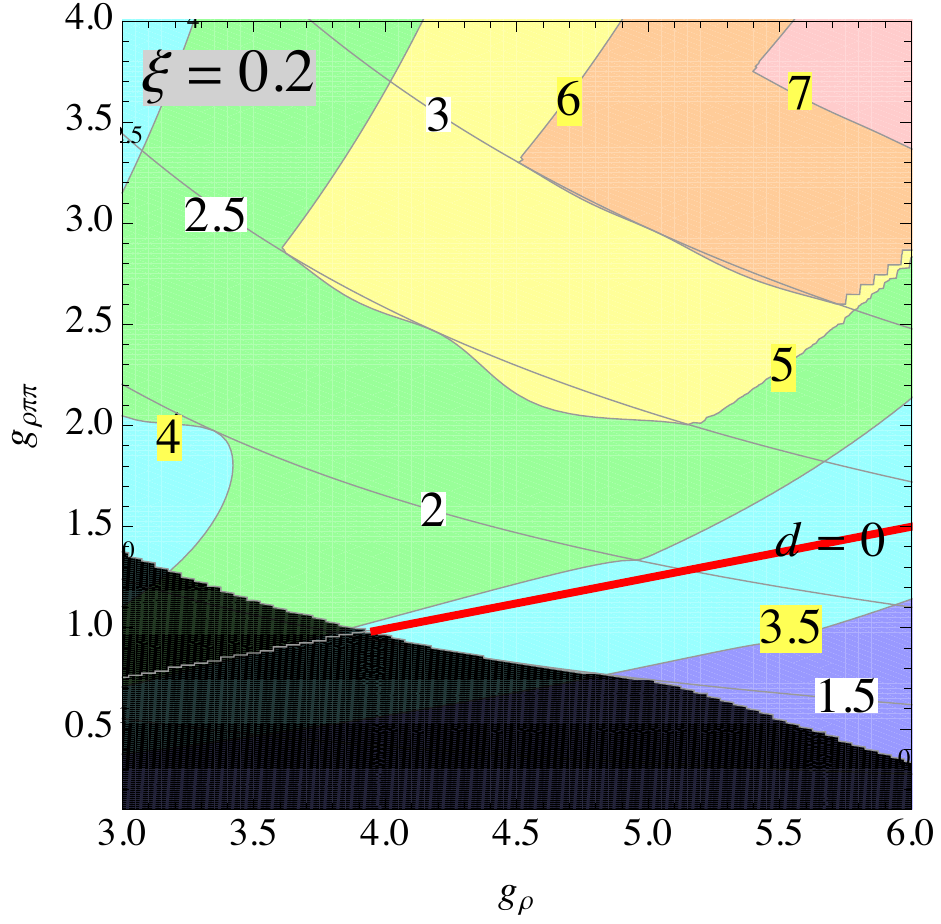}
\caption{LHC exclusion limits for $\xi=0.2$.}
\label{limitSO5xi02}
\end{minipage}
\hspace{0.5cm}
\begin{minipage}[b]{0.47\linewidth}
\centering
\includegraphics[width=70 mm]{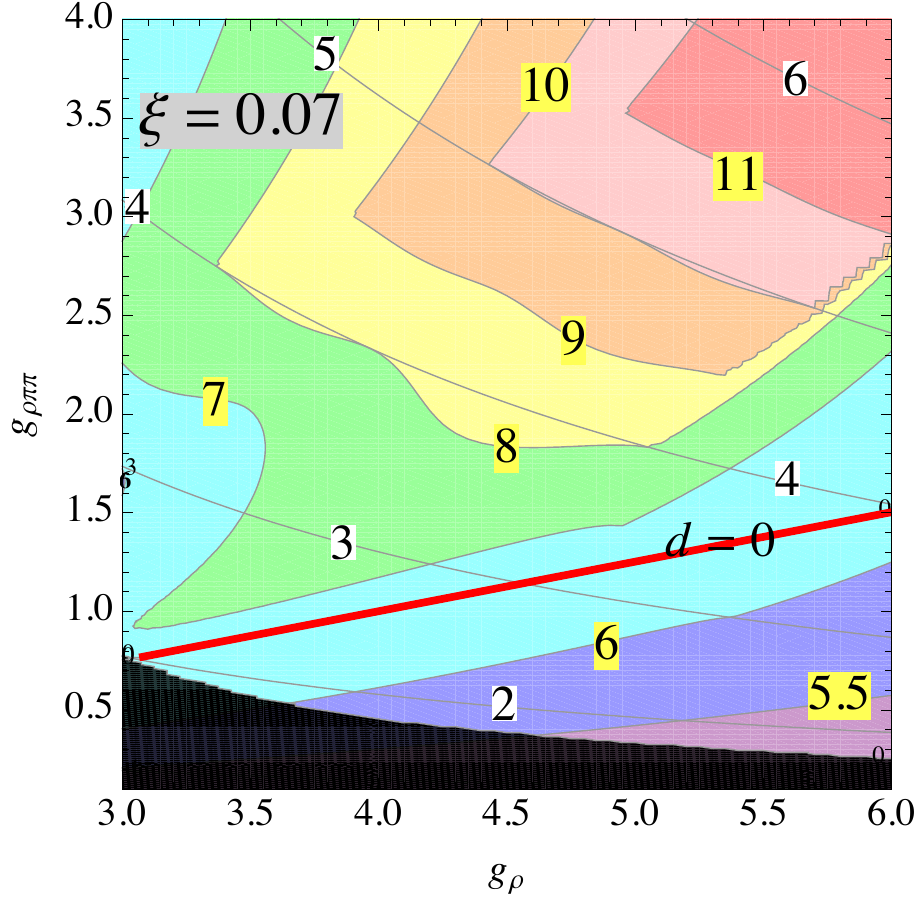}
\caption{LHC exclusion limits for $\xi=0.07$.}
\label{limitSO5xi007}
\end{minipage}
\end{figure}

In this section we will focus on the minimal $SO(5)\rightarrow SO(4)$ composite Higgs model with lightest vector resonances $\rho_{L}$ and $\rho_{R}$ corresponding to $SO(4)_{hidden}$. This setup should be representative for the LHC phenomenology of spin-1 resonances in general composite Higgs models. The most general leading order effective Lagrangian related to $SO(5)\times SO(4)_{hidden}\rightarrow SO(4)$ has 3 free parameters, which can be chosen as $\xi,\ g_{\rho}$ and $g_{\rho\pi\pi}$, where $g_{\rho}$ is the 'hidden' gauge coupling and $g_{\rho\pi\pi}$ describes the mean interaction of a single $\rho$ resonance with two pions corresponding to $W_{L}$ and $Z_{L}$. The mass of spin-1 resonances is given by $m_{\rho}^{2}\approx2g_{\rho}g_{\rho\pi\pi} v^2/\xi$. Interactions of heavy spin-1 eigenstates with SM matter fields are introduced by the admixture of heavy $\tilde{\rho}_{L}$ and $\tilde{\rho}_{R}$ eigenstates in Standard Model fields
\bea
W^{\pm}_{\mu} & \approx & \tilde{W}^{\pm}_{\mu}-\frac{\sqrt{2}}{2}\sqrt{2-\xi}\; \frac{g}{g_{\rho}}\tilde{\rho}^{\pm}_{L\; \mu} \nn
Z_{\mu} & \approx & \tilde{Z}_{\mu}-\frac{\sqrt{2-\xi}}{\sqrt{2}}\frac{g^2-g'^2}{g_{\rho}\sqrt{g^2+g'^2}}\tilde{\rho}_{L\;\mu}^{0}-\frac{2\sqrt{2-2\xi}}{\left( 2-\xi\right) ^{3/2}}\frac{g'^2}{g_{\rho}\sqrt{g^2+g'^2}}\tilde{\rho}_{R\;\mu}^{0} \nn
A_{\mu} & \approx & \tilde{A}_{\mu}-\sqrt{\frac{2}{2-\xi}}\frac{e}{g_{\rho}}\tilde{\rho}_{L\;\mu}+\sqrt{\frac{2-2\xi}{2-\xi}}\frac{e}{g_{\rho}}\tilde{\rho}_{R\;\mu}.
\eea
The couplings of $\rho$ resonances with the third generation of fermions can be enhanced by partial compositeness, but we neglect this effect here. One can notice that the coupling of a $\rho$ resonance to two fermions is enhanced for small values of $\xi$, while for a given value of $m_{\rho}$ and $g_{\rho}$ the coupling of a $\rho$ resonance to two SM gauge bosons scales down with $\xi$. Numerical analysis shows that spin-1 resonances are mainly produced in the Drell-Yan process. The production cross section of neutral (smooth line) and charged (dashed line) $\rho_{L}$ resonances at the LHC at $8\tev$ and $14\tev$ is given in Fig. \ref{DYg4SO5xi02} for $\xi=0.2$ and $g_{\rho}=4$. Branching ratios of a neutral resonance with these parameters are given in Fig. \ref{BRv0g4SO5xi02}. One can notice that while the width is dominated by the production of SM gauge boson pairs or a Higgs boson accompanied by a gauge boson, branching ratios into quark pairs and lepton pairs are non-negligible. As it turns out, due to low backgrounds, LHC searches for dilepton resonances \cite{:2012vb} give at present the most stringent exclusion limits for spin-1 resonances. The impact of LHC searches on the parameter space of spin-1 resonances spanned by $g_{\rho}$ and $g_{\rho\pi\pi}$ can be seen in Fig. \ref{limitSO5xi02} for $\xi=0.2$ and in Fig. \ref{limitSO5xi007} for $\xi=0.07$. The various colors in these figures describe the approximate cutoff of the effective description derived by the analysis of perturbative unitarity of longitudinal gauge boson scattering (numbers on yellow background give the energy of the cutoff in $\tev$). Black curves and numbers on white background correspond to the $\rho$ resonance mass, also in $\tev$. White regions are excluded because the cutoff drops below the resonance mass. Black regions are excluded by direct LHC searches. One can notice that the exclusion limits go up to $1.5\tev$ for $\xi=0.2$ and $2\tev$ for $\xi=0.07$. They go down with growing values of $g_{\rho}$, as the interaction with SM matter fields become suppressed.

\begin{figure}[h!]
\begin{minipage}[b]{0.47\linewidth}
\centering
\includegraphics[width=70 mm]{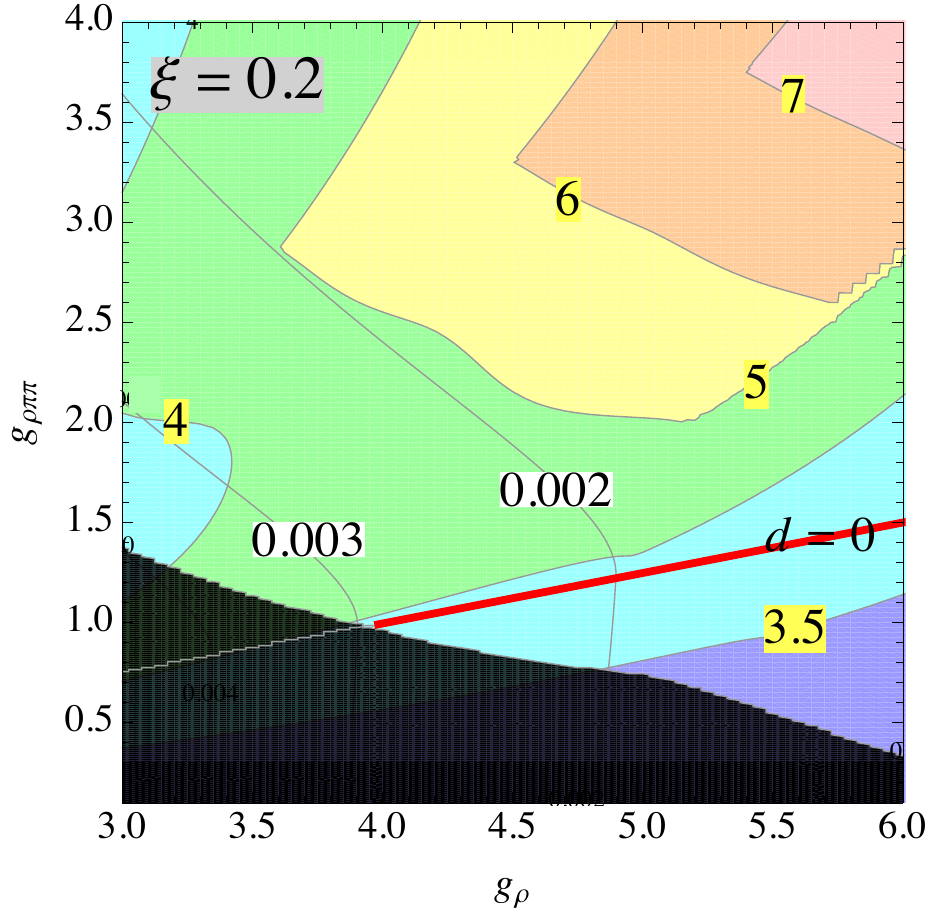}
\caption{Values of $\hat{S}$ for $\xi=0.2$.}
\label{ShatSO5xi02}
\end{minipage}
\hspace{0.5cm}
\begin{minipage}[b]{0.47\linewidth}
\centering
\includegraphics[width=70 mm]{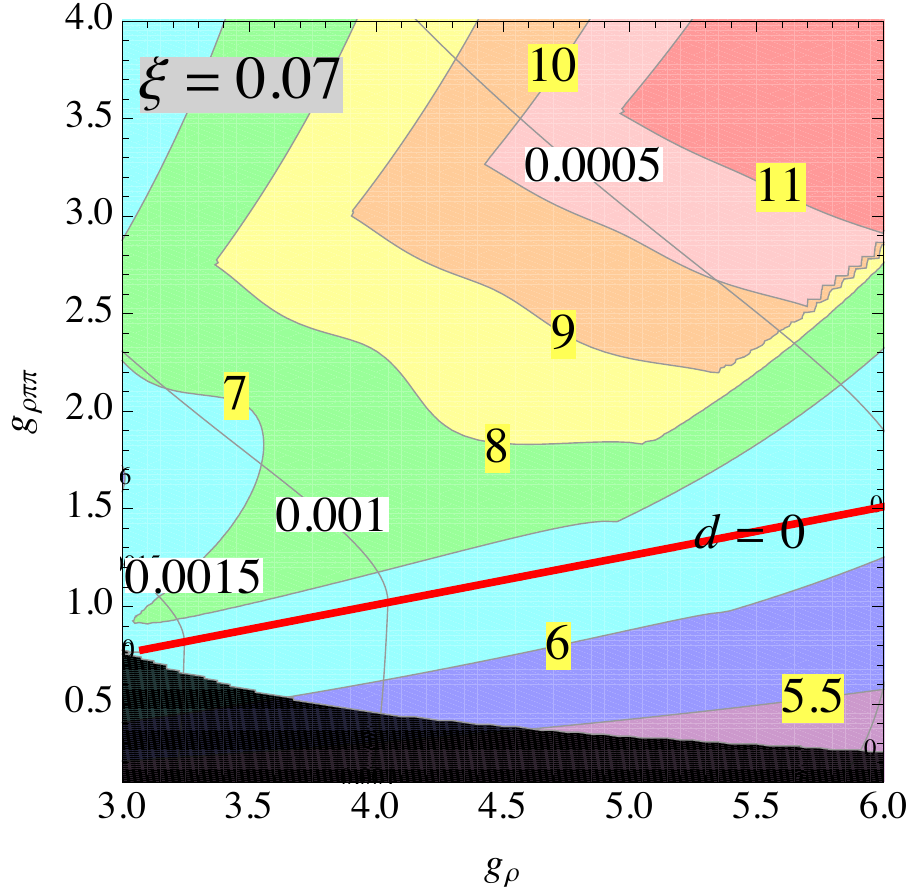}
\caption{Values of $\hat{S}$ for $\xi=0.07$.}
\label{ShatSO5xi007}
\end{minipage}
\end{figure}

It is interesting to compare exclusion limits from direct searches with the ones originating form electroweak precision tests. Figures \ref{ShatSO5xi02} and \ref{ShatSO5xi007} show the $\hat{S}=\frac{g^2}{16\pi}S$ parameter for $\xi=0.2$ and $\xi=0.07$ (black curves with labels on white background). It can be seen that the LHC direct searches are already exploring the parameter space consistent with electroweak precision data.

\section{Conclusions}

Properties and LHC phenomenology of spin-1 resonances were studied using a simple, general effective description built using the hidden local symmetry formalism. It shows that for small values of $\xi$ the spin-1 resonance coupling to two fermions is enhanced. The resonances are mainly Drell-Yan produced and they decay mostly to SM gauge boson pairs or a Higgs boson accompanied by a gauge boson, but branching ratios into fermion pairs are non-negligible. The most stringent exclusion limits are obtained from searches for resonant dilepton production at the LHC. Direct searches are already probing the parameter space of spin-1 resonances allowed by electroweak precision data.


\section*{Acknowledgments}
This work has been supported by a Marie Curie Initial Training Network Fellowship PITN-GA-2009-237920-UNILHC.

\end{document}